\documentclass[a4paper]{article}

\usepackage{INTERSPEECH2020}
\usepackage{subfigure}

\usepackage{amssymb}
\usepackage[table]{xcolor}
\usepackage{multirow}
\usepackage{graphicx}
\usepackage{tabularx}
\usepackage{hyperref}
\usepackage[ruled,vlined]{algorithm2e}

\newlength\mylength
\begin{document}

\title{Unsupervised Audio Source Separation using Generative Priors}
\name{Vivek Narayanaswamy$^1$\thanks{This work was supported in part by the ASU SenSIP Center, Arizona State University. This work was performed under the auspices of the U.S. Department of Energy by Lawrence Livermore National Laboratory under Contract DE-AC52-07NA27344.}, Jayaraman J. Thiagarajan$^2$, Rushil Anirudh$^2$ and Andreas Spanias$^1$}
\address{
	$^1$SenSIP Center, School of ECEE, Arizona State University, Tempe, AZ\\
	$^2$Lawrence Livermore National Labs, 7000 East Avenue, Livermore, CA}
\email{vnaray29@asu.edu, jjayaram@llnl.gov, anirudh1@llnl.gov, spanias@asu.edu}

%

\maketitle

\begin{abstract}
  State-of-the-art under-determined audio source separation systems rely on supervised end-end training of carefully tailored neural network architectures operating either in the time or the spectral domain. However, these methods are severely challenged in terms of requiring access to expensive source level labeled data and being specific to a given set of sources and the mixing process, which demands complete re-training when those assumptions change. This strongly emphasizes the need for unsupervised methods that can leverage the recent advances in data-driven modeling, and compensate for the lack of labeled data through meaningful priors. To this end, we propose a novel approach for audio source separation based on generative priors trained on individual sources. Through the use of projected gradient descent optimization, our approach simultaneously searches in the source-specific latent spaces to effectively recover the constituent sources. Though the generative priors can be defined in the time domain directly, e.g. WaveGAN, we find that using spectral domain loss functions for our optimization leads to good-quality source estimates. Our empirical studies on standard spoken digit and instrument datasets clearly demonstrate the effectiveness of our approach over classical as well as state-of-the-art unsupervised baselines.
    
\end{abstract}

\section{Introduction}
Audio source separation, the process of recovering constituent source signals from a given audio mixture, is a key component in downstream applications such as audio enhancement and music information retrieval \cite{spanias2006audio, spanias2015advances}. Typically formulated as an inverse optimization problem, source separation has been traditionally solved using a broad class of matrix factorization methods~\cite{makino,karhunen1995nonlinear,thiagarajan2013mixing}, e.g., Independent Component Analysis (ICA) and Principal Component Analysis (PCA). While these methods are known to be effective in over-determined scenarios, i.e. the number of mixture observations is greater than the number of sources, they are severely challenged in under-determined settings~\cite{wang2016over}. Consequently, in the recent years, supervised deep learning based solutions have become popular for under-determined source separation~\cite{stoller2018wave, luo2019conv, lluis2018end, takahashi2018mmdenselstm, grais2018raw, defossez2019demucs}. These approaches can be broadly classified into time domain and spectral domain methods, and often produce  state-of-the-art performance on standard benchmarks. Despite their effectiveness, there is a fundamental drawback with supervised methods. In addition to requiring access to large number of observations, a supervised source separation model is highly specific to the given set of sources and the mixing process, consequently requiring complete re-training when those assumptions change. This motivates a strong need for the next generation of unsupervised separation methods that can leverage the recent advances in data-driven modeling, and compensate for the lack of labeled data through meaningful priors.

Utilizing appropriate priors for the unknown sources has been an effective approach to regularize the ill-conditioned nature of source separation. Examples include non-Gaussianity, statistical independence, and sparsity~\cite{virtanen2003sound}. With the emergence of deep learning methods, it has been shown that choice of the network architecture implicitly induces a structural prior for solving inverse problems~\cite{ulyanov2018deep}. Based on this finding, Tian \textit{et al.} recently introduced a \textit{deep audio prior} (DAP)~\cite{tian2019deep} that directly utilizes the structure of a randomly initialized neural network to learn time-frequency masks that isolate the individual components in the mixture audio without any pre-training. Interestingly, DAP was shown to outperform several classical priors.

\noindent Here, we consider an alternative approach for under-determined source separation based on \textit{data priors} defined via deep generative models, and in particular using generative adversarial networks (GANs)~\cite{goodfellow2014generative}. We hypothesize that such a data prior will produce higher quality source estimates by enforcing the estimated solutions to belong to the data manifold. While GAN priors have been successfully utilized in inverse imaging problems~\cite{bora2017compressed, zhu2017unpaired, shah2018solving, anirudh2020mimicgan} such as denoising, deblurring, compressed recovery etc., their use in source separation has not been studied yet -- particularly in the context of audio. In this paper, we propose a novel unsupervised approach for source separation that utilizes multiple source-specific priors and employs \textit{Projected Gradient Descent} (PGD)-style optimization with carefully designed spectral-domain loss functions. Since our approach is an inference-time technique, it is extremely flexible and general such that it can be used even with a single mixture. We utilize the time-domain based WaveGAN~\cite{donahue2018adversarial} model to construct the source-specific priors, and interestingly, we find that using spectral losses for the inversion leads to superior quality results. Using standard benchmark datasets (spoken digit audio (SC09), drums and piano), we evaluate the proposed approach under the assumption that mixing process is known. From our rigorous empirical study, we find that the proposed \textit{data prior} is consistently superior to other commonly adopted priors, including the recent deep audio prior~\cite{tian2019deep}. The code for this work can be found at \url{https://github.com/vivsivaraman/sourcesepganprior}.

\section{Designing Priors for Inverse Problems}
Despite the advances in learning methods for audio processing, under-determined source separation remains a critical challenge. Formally, in our setup, the number of mixtures or observations $m \ll n$, \textit{i.e.} the number of sources. A common approach to make this ill-defined problem tractable is to place appropriate priors to restrict the solution space. Existing approaches can be broadly classified into the following categories:

\noindent \textbf{(i) Statistical Priors.} This includes the class of matrix factorization methods conventionally used in source separation. For example in ICA, we enforce the assumptions of non-Gaussianity as well as statistical independence  between the sources. On the other hand, PCA enforces statistical independence between the sources by linear projection onto mutually orthogonal subspaces. KernelPCA~\cite{mika1999kernel} induces the same prior in a reproducing kernel Hilbert space. Another popular approach is Non-negative matrix factorization (NMF), which places a non-negativity prior on the estimated basis matrices \cite{fevotte2018single}. Finally, a sparsity prior ($\ell_1$)~\cite{virtanen2003sound} placed either in the observed domain or in the expansion via an appropriate basis set or a dictionary has also been widely adopted to regularize this problem.

\noindent \textbf{(ii) Structural Priors.} Recent advances in deep neural network design have shown that certain carefully chosen networks have the innate capability to effectively regularize or behave as a prior to solve ill-posed inverse problems. These networks essentially capture the underlying statistics of data, independent of the task-specific training. These \textit{structural priors} have produced state-of-the-art performance in inverse imaging problems~\cite{ulyanov2018deep} and recently, Tian \textit{et al.}~\cite{tian2019deep} utilized the structure of an U-Net \cite{ronneberger2015u} model to learn time-frequency masks that can isolate the individual components in the mixture audio.

\noindent \textbf{(iii) GAN Priors.} A third class of methods have relied on priors defined via generative models, e.g. GANs~\cite{goodfellow2014generative}. GANs can learn parameterized non-linear distributions $p(X;\mathbf{z})$ from a sufficient amount of unlabeled data $X$~\cite{donahue2018adversarial,radford2015unsupervised}, where $\mathbf{z}$ denotes the latent variables of the model. In addition to readily sampling from trained GAN models, they can be leveraged as an effective prior for $X$. Popularly referred to as \textit{GAN priors}, they have been found to be highly effective in challenging inverse problems \cite{shah2018solving,anirudh2020mimicgan}. In its most general form, when one attempts to recover the original data $\mathbf{x}$ from its corrupted version $\mathbf{\tilde{x}}$ (observed), one can maximize the posterior distribution $p(X = \mathbf{x} | \mathbf{\tilde{x}} ; \mathbf{z} )$ by searching in the latent space of a pre-trained GAN.  
Since this posterior distribution cannot be expressed analytically, in practice, we utilize an iterative approach such as \textit{Projected Gradient Descent} (PGD) to estimate the latent features $\hat{\mathbf{z}}$ followed by sampling from the generator, \textit{i.e} $p(X ; \mathbf{z} = \hat{\mathbf{z}} )$. 

\noindent \textbf{Proposed Work.} In this work, we propose to utilize GAN priors to solve the problem of under-determined source separation. Existing solutions with data priors utilize a single GAN model to perform the inversion process~\cite{anirudh2020mimicgan}. However, by design, source separation requires the simultaneous estimation of multiple disparate source signals. While one can potentially build a generative model that can jointly characterize all sources, it will require significantly large amounts of data. Hence, we advocate the use of source-specific generative models and generalizing the PGD optimization with multiple GAN priors. In addition to reducing the data needs, this approach provides the crucial flexibility of handling new sources, without the need for retraining the generative models for all sources. From our study, we find that utilizing multiple GAN priors $\{\mathcal{G}_i | i =1 \dots K\}$  to be highly effective for under-determined source separation. In particular, we choose a popular waveform synthesis model WaveGAN \cite{donahue2018adversarial} as our GAN prior $\mathcal{G}_i$ as we found the generated samples to be of high perceptual quality. While we utilize time domain GAN prior models, we find that spectral domain loss functions are critical in source estimation using PGD.  

\section{Approach}
	

\begin{figure}
	\centering
	\includegraphics[width=\columnwidth]{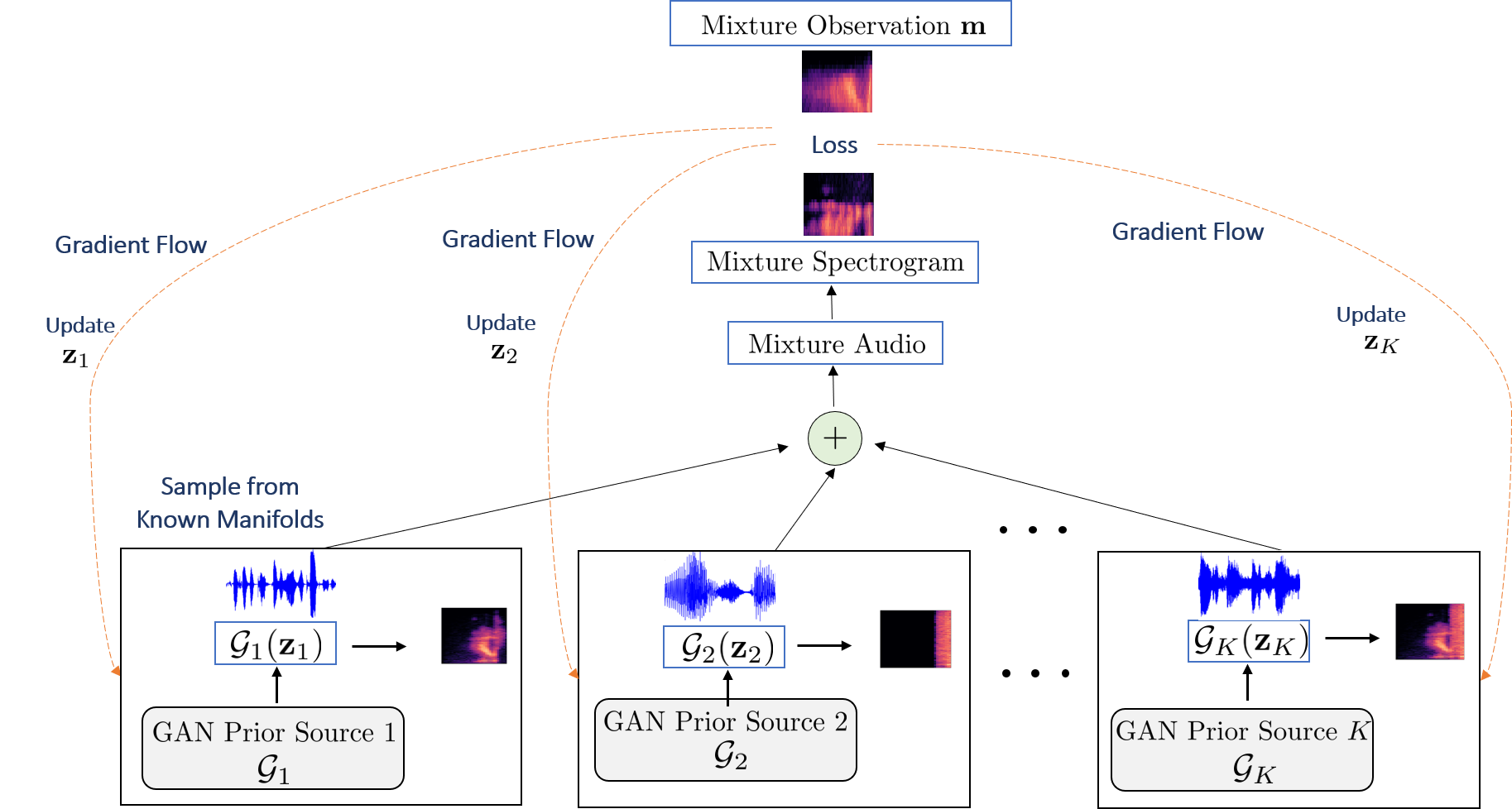}
	
	\caption{An overview of the proposed unsupervised source separation system. }
	\vspace{-0.1in}
	\label{fig:approach}
\end{figure}

Audio source separation involves the process of recovering constituent sources \{$\mathbf{s}_{i} \in \mathbb{R}^{d} | i = 1\cdots K$\} from a given audio mixture $\mathbf{m} \in \mathbb{R}^{d}$,  where $K$ is the total number of sources and $d$ is the number of time steps. In this paper, without loss of generality, we assume the source and mixtures to be mono-channel and the mixing process to be a sum of sources i.e., $\mathbf{m} = \sum_{i = 1}^{K} \mathbf{s}_i$. Figure \ref{fig:approach} provides an overview of our proposed approach for unsupervised source separation. Here, we sample the source audio from the respective priors and perform additive mixing to reconstruct the mixture \textit{i.e},  $\hat{\mathbf{m}} = \sum_{i=1}^{K}\mathcal{G}_i(\mathbf{z}_i)$. The mixture is then processed to obtain the corresponding spectrogram. In addition, we also compute the source level spectrograms. We perform source separation by efficiently searching the latent space of the source-specific priors $\mathcal{G}_i$ using  \textit{Projected Gradient Descent} optimizing a spectral domain loss function $\mathcal{L}$. More formally, for a single mixture $\mathbf{m}$, our objective function is given by,
\begin{equation}
\label{generalinv}
\{{\mathbf{z}}^{*}_i\}_{i=1}^{K} =arg\min_{\mathbf{z}_1,\mathbf{z}_2 \dots \mathbf{z}_K} \mathcal{L}(\hat{\mathbf{m}}, \mathbf{m})  +  \mathcal{R}(\{\mathcal{G}_{i}(\mathbf{z}_i)\}),
\end{equation}
where the first term measures the discrepancy between the true and estimated mixtures and the second term is an optional regularizer on the estimated sources. In every PGD iteration, we perform a projection $\mathcal{P}$, where we constrain the $\{{\mathbf{z}}_i\}_{i=1}^{K}$ to their respective manifolds.  Upon completion of this optimization, the sources can be obtained as $
\hat{\mathbf{s}}^{*}_i = \mathcal{G}_i(\mathbf{z}_i^{*}), \forall i$.
Here, we reformulate the process of source separation by first estimating the source-specific latent features $\mathbf{z}_i^{*}$ followed by sampling from the respective generators. There are two key ingredients that are critical to the performance of our approach: (i) choice of a good quality \textit{GAN Prior} for every source and (ii) carefully chosen loss functions to drive the PGD optimization. We now elaborate our methodology in the rest of this section.

\begin{algorithm}[t]
	\SetAlgoLined
	\textbf{Input}: \hspace{0.05in} Unlabeled mixture $\mathbf{m}$, No. of sources $K$, \\
	 \hspace{0.45in} Pre-trained \textit{GAN Priors} $\{ \mathcal{G}_i\}_{i =1 \dots K}$ 
	 
	 \textbf{Output}: Estimated sources $\{ \hat{\mathbf{s}}^{*}_i\}_{i =1 \dots K}$ 
	 
	 \textbf{Initialization}: $\{\hat{\mathbf{z}}_i\}_{i =1 \dots K} = \mathbf{0} \in \mathbb{R}^{d_z}$ 
	 	
	\For{$t\gets1$ \KwTo $T$}{
		$\hat{\mathbf{m}} = \sum_{i=1}^{K}\mathcal{G}_i(\hat{\mathbf{z}}_i)$\\
		Compute source level and mixture spectrograms\\
		Compute loss $\mathcal{L}$ using \ref{totalloss}\\
		$\hat{\mathbf{z}}_i\gets\hat{\mathbf{z}}_i -\eta\nabla_{z}(\mathcal{L}) \hspace{0.5in} \forall i = 1 \dots K$\\
		$\hat{\mathbf{z}}_i\gets\mathcal{P}(\hat{\mathbf{z}}_i) \hspace{0.1in}$ $\mathcal{P}$ projects $\{\mathbf{z}_i\}_{i =1 \dots K}$ onto  the \\
		\hspace{0.7in} manifold, i.e., clipped to $[-1, 1]$	
}
return $\{ \hat{\mathbf{s}}^{*}_i\} =  \mathcal{G}_i(\mathbf{z}_i^{*}), \forall i$
\caption{Proposed Approach.}
\label{pgdprocedure}
\end{algorithm}

\subsection{WaveGAN for Data Prior Construction}
WaveGAN \cite{donahue2018adversarial} is a popular generative model capable of synthesizing raw waveform audio. It has exhibited success in producing audio from different domains such as speech and musical instruments. Both the generator and discriminator of the WaveGAN model are similar in construction to DCGAN \cite{radford2015unsupervised} with certain architectural changes to support audio generation. The generator $\mathcal{G}$ transforms the latent features $\mathbf{z} \in \mathbb{R}^{d_z} $ where $d_z = 100$ from a uniform distribution in $[-1,1]$,  to produce waveform audio $\mathcal{G}(\mathbf{z})$ of dimension $d$ = 16384 which is approximately of 1s duration at a sampling rate of 16kHz. The discriminator $\mathcal{D}$ regularized using phase shuffle learns to distinguish between the real and synthesized samples. The WaveGAN is trained to optimize the Wasserstein loss with gradient penalty (WGAN-GP) as prescribed in \cite{gulrajani2017improved}. 

\begin{table}[t]
	\centering
	\renewcommand*{\arraystretch}{1.2}
	\caption{Performance  metrics averaged across $1000$ cases for the Digit-Piano ($K$ = 2) experiment (While higher Spectral SNR and SIR are better, lower RMS Env.Distance is better).}
	\resizebox{\columnwidth}{!}{%
		\begin{tabular}{|c|cc|cc|cc|}
			\hline
			\multirow{2}{*}{\textbf{Method}} &
			\multicolumn{2}{c|}{\textbf{Spectral SNR (dB)}} &
			\multicolumn{2}{c|}{\textbf{RMS Env. Distance}} &
			\multicolumn{2}{c|}{\textbf{SIR (dB)}} \\ \cline{2-7} 
			&
			\textbf{Digit} &
			\textbf{Piano} &
			\textbf{Digit} &
			\textbf{Piano} &
			\textbf{Digit} &
			\textbf{Piano} \\ \hline \hline
			{FastICA}    & -2.13 & -13.45 & 0.22 & 0.61 & -4.12 & -0.66 \\
			{PCA}        & -2.04 & -12.01 & 0.22 & 0.54 & -4.13 & -1.44 \\
			{Kernel PCA} & -2.04 & -3.30  & 0.22 & 0.26 & -4.13 & -1.61 \\
			{NMF}        & -2.21 & -5.80  & 0.23 & 0.26 & -4.09 & 2.53  \\
			{DAP}        & -1.77 & \textbf{2.72}   & 0.22 & 0.22 & 2.20  & -3.10 \\
			\hline \hline
			{Proposed} &
			\textbf{1.06} &
			\textbf{2.73} &
			\textbf{0.17} &
			\textbf{0.21} &
			\textbf{3.91} &
			\textbf{8.57} \\ \hline
		\end{tabular}%
	}
	\label{digit_piano}
\end{table}

\begin{table}[t]
	\centering
	\renewcommand*{\arraystretch}{1.2}
	\caption{Performance  metrics averaged across $1000$ cases for the Drums-Piano ($K$ = 2) experiment.}
	\resizebox{\columnwidth}{!}{%
		\begin{tabular}{|c|cc|cc|cc|}
			\hline
			\multirow{2}{*}{\textbf{Method}} &
			\multicolumn{2}{c|}{\textbf{Spectral SNR (dB)}} &
			\multicolumn{2}{c|}{\textbf{RMS Env. Distance}} &
			\multicolumn{2}{c|}{\textbf{SIR (dB)}} \\ \cline{2-7}  
			& \textbf{Drums} & \textbf{Piano} & \textbf{Drums} & \textbf{Piano} & \textbf{Drums} & \textbf{Piano} \\ \hline \hline
			{FastICA}    & -5.25          & -13.52         & 0.24           & 0.61           & -6.51          & -1.45          \\
			{PCA}        & -5.19          & -12.33         & 0.24           & 0.56           & -6.53          & -2.69          \\
			{Kernel PCA} & -5.19          & -3.36          & 0.24           & 0.25           & -6.53          & -2.02          \\
			{NMF}        & -5.39          & -5.84          & 0.24           & 0.26           & -6.59          & 3.84           \\
			{DAP}        & -4.20          & 2.97           & 0.22           & \textbf{0.21 }          & -21.62         & \textbf{11.22} \\ \hline \hline
			{Proposed}   & \textbf{0.84}  & \textbf{3.06}  & \textbf{0.10}  & \textbf{0.21}  & \textbf{11.70} & {9.80}  \\ \hline
		\end{tabular}%
	}
	\label{drums_piano}
\end{table}

\begin{table}[!t]
	\centering
	\renewcommand*{\arraystretch}{1.2}
	\caption{Performance  metrics averaged across $1000$ cases for the Digit-Drums ($K$ = 2) experiment.}
	\resizebox{\columnwidth}{!}{%
		\begin{tabular}{|c|cc|cc|cc|}
			\hline
			\multirow{2}{*}{\textbf{Method}} &
			\multicolumn{2}{c|}{\textbf{Spectral SNR (dB)}} &
			\multicolumn{2}{c|}{\textbf{RMS Env. Distance}} &
			\multicolumn{2}{c|}{\textbf{SIR (dB)}} \\ \cline{2-7}  
			& \textbf{Digit} & \textbf{Drums} & \textbf{Digit} & \textbf{Drums} & \textbf{Digit} & \textbf{Drums} \\ \hline \hline
			{FastICA}    & 2.91           & -21.01         & \textbf{0.13  }         & {0.82}  & 3.10           & 0.09           \\
			{PCA}        & 2.99           & -20.00         & \textbf{0.13 }          & 0.77           & 3.12           & 0.02           \\
			{Kernel PCA} & 2.99           & -10.53         & \textbf{0.13 }          & 0.35           & 3.12           & 0.85           \\
			{NMF}        & 3.01           & -13.75         & \textbf{0.13   }       & 0.39           & 3.20           & -0.98          \\
			{DAP}        & \textbf{3.59}  & \textbf{0.92}  & 0.14           & 0.14           & 4.24           & -11.48         \\ \hline \hline
			{Proposed}   & {2.32}  & {0.42}  & {0.15}  & \textbf{0.10}  & \textbf{25.91} & \textbf{23.68} \\ \cline{1-7}
		\end{tabular}%
	}
	\label{digit_drums}
\end{table}

Given the ability of WaveGAN to synthesize high quality audio, the pre-trained generator of WaveGAN was used to define the \textit{GAN Prior}. In our formulation, instead of using a single \textit{GAN Prior} trained jointly for all sources, we construct $K$ independent source-specific priors. 

\subsection{Losses}
In order to obtain high-quality source estimates using GAN priors, we propose a novel yet intuitive combination of spectral-domain losses. Though one can utilize time-domain metrics such as the Mean-Squared Error (MSE) to compare the observed and synthesized mixtures, we find that even small variations in the phases of sources estimated from our priors can lead to higher error values. This in turn can misguide the PGD optimization process and may lead to poor convergence. This corroborates with the findings in \cite{defossez2018sing}.


\subsubsection{Multiresolution Spectral Loss ($\mathcal{L}_{ms}$)}
This loss term measures the $\ell_1$-norm between log magnitudes of the reconstructed spectrogram and the input spectrogram at $L$ spatial resolutions. This is used to enforce perceptual closeness between the two mixtures at varying spatial resolutions. Denoting $\mathbf{m}$ as the input mixture and $\hat{\mathbf{m}}$ as the estimated mixture, the loss $\mathcal{L}_{ms}$ is defined as
\begin{align}
\nonumber \mathcal{L}_{ms}= \sum_{l=1}^{L} & \bigg\| \log(1+|STFT^{l}(\mathbf{m})|^2)  \\	
& -	\log(1+|STFT^{l}(\hat{\mathbf{m}})|^2)\bigg\|_1,
\end{align}where $|STFT^{l}(.)|$ represents the magnitude spectrograms at the $l^{th}$ spatial resolution and $L = 3$. We compute the magnitude spectrogram at different resolutions by performing a simple average pooling operation with bilinear interpolation.  

\subsubsection{Source Dissociation Loss ($\mathcal{L}_{sd}$)}
Minimizing this loss, defined as the aggregated gradient similarity between the spectrograms of the estimated sources, enforces them to be systematically different. Similar to \cite{tian2019deep, zhang2018single}, we define this as a product of the normalized gradient fields of the log magnitude spectrograms computed at $L$ spatial resolutions. In the case where there are $K$ constituent sources, we compute this between every pair of sources. Formally,
\begin{align}
\nonumber \mathcal{L}_{sd} = \sum_{i=1}^{K}\sum_{j=i+1}^{K}\sum_{l=1}^{L} & ||\Psi(\log(1+|STFT^{l}(\mathcal{G}_{i}(\hat{\mathbf{z}}_i))|^2), \\  	
&\log(1+|STFT^{l}(\mathcal{G}_{j}(\hat{\mathbf{z}}_j))|^2))||_F,
\end{align}
where $\Psi(x, y)$ = $tanh(\lambda_1|\nabla x|) \odot tanh(\lambda_2|\nabla y|)$. ($\odot$ represents element-wise multiplication) and $L = 3$. 
The weights $\lambda_1$ and $\lambda_2$ are set at $\lambda_1 = \frac{\sqrt{|\nabla y|_F}}{\sqrt{|\nabla x|_F}}$ and $\lambda_2 = \frac{\sqrt{|\nabla x|_F}}{\sqrt{|\nabla y|_F}}$.

\subsubsection{Mixture Coherence Loss ($\mathcal{L}_{mc}$)}
Along with $\mathcal{L}_{ms}$, this loss, defined using gradient similarity between original and reconstructed mixtures, ensures that our PGD optimization produces meaningful reconstructions:
\begin{align}
\nonumber \mathcal{L}_{mc} = -\sum_{l=1}^{L} & ||\Psi(\log(1+|STFT^{l}(\mathbf{m})|^2), \\  	
&\log(1+|STFT^{l}(\hat{\mathbf{m}}))|^2))||_F
\end{align}

%
%

\subsubsection{Frequency Consistency Loss ($\mathcal{L}_{fc}$)}
\begin{figure}
	\centering
	\includegraphics[width=\columnwidth]{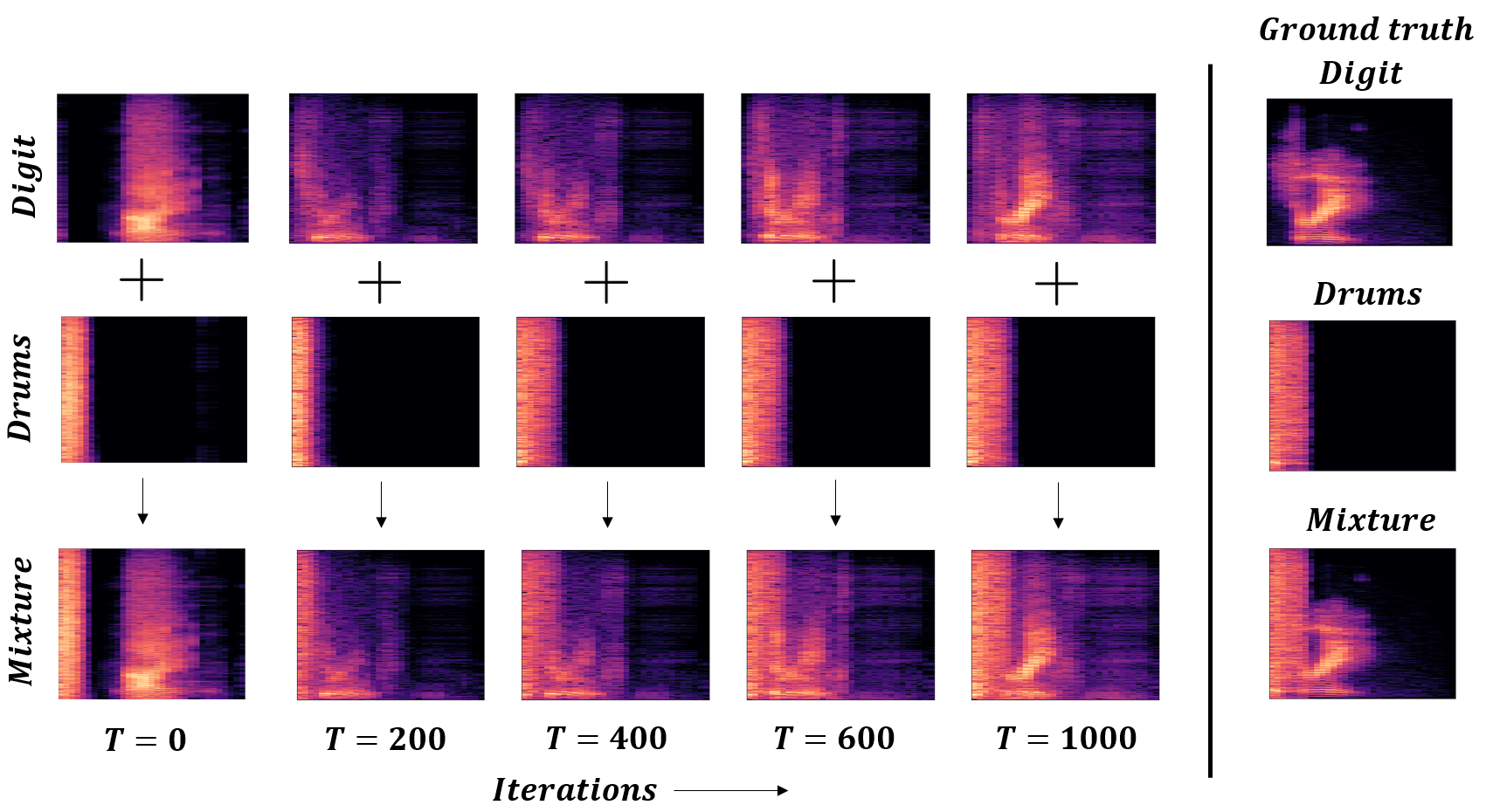}
	
	\caption{Demonstration of our proposed approach using a digit-drum example. Through the use of multiple \textit{GAN Priors} $\mathcal{G}_i$, our algorithm efficiently searches the source-specific latent spaces to estimate the underlying sources.}
	\vspace{-0.1in}
	\label{fig:evolution}
\end{figure}

This loss helps improve perceptual similarity between the magnitude spectrograms of the input and synthesized mixtures by constraining components within a particular temporal bin of the spectrograms to remain consistent over the entire frequency range, i.e.,
\begin{equation}
\mathcal{L}_{fc} = \sum_{t=1}^{T} \sum_{f=1}^{F} \frac{\log(1+|STFT(\mathbf{m})[t,f]|)}{\log(1+|STFT(\hat{\mathbf{m}})[t,f]|)}.
\end{equation}
The overall loss function for our source separation algorithm is thus obtained as:
\begin{equation}
\mathcal{L} = \beta_1 \mathcal{L}_{ms} +\beta_2  \mathcal{L}_{sd}+ \beta_3 \mathcal{L}_{mc} + \beta_4 \mathcal{L}_{fc}
\label{totalloss}
\end{equation}Through hyperparameter search we identified that $\beta_1 =0.8, \beta_2=0.3, \beta_3 = 0.1, \beta_4 = 0.4$ to be effective in our experiments. Note, in our computations we obtain the spectrograms by computing the Short Time Fourier Transform (STFT) on the waveform in frames of length 256, hop size of 128 and FFT length of 256. The procedure for our approach is showed in Algorithm \ref{pgdprocedure}. Figure \ref{fig:evolution} illustrates the progressive estimation of the unknown sources using our approach.

\begin{table}[t]
	\centering
	\renewcommand*{\arraystretch}{1.2}
	\caption{Performance  metrics averaged across $1000$ cases for the Digit-Drums-Piano ($K$ = 3) experiment.}
	\resizebox{\columnwidth}{!}{%
		\label{digit_drums_piano}
		\begin{tabular}{|c|c|c|c|c|c|c|}
			\hline
			\textbf{Metric} & \textbf{Source} & \textbf{FastICA} & \textbf{PCA} & \textbf{Kernel PCA} & \textbf{NMF} & \textbf{Proposed} \\ \hline \hline
			\multirow{3}{*}{Spectral SNR (dB)} & Digit & -2.95 & -2.47  & -2.47 & -2.47         & \textbf{0.77}  \\
			& Drums & -10.8 & -19.81 & -8.1  & -12.84        & \textbf{0.64}  \\
			& Piano & 0.27  & 0.1    & -0.94 & \textbf{4.94} & 2.64           \\
			\hline \hline
			\multirow{3}{*}{RMS Env. Distance} & Digit & 0.24  & 0.23   & 0.23  & 0.23          & \textbf{0.17}  \\
			& Drums & 0.4   & 0.75   & 0.28  & 0.37          & \textbf{0.1}   \\
			& Piano & 0.23  & 0.31   & 0.25  & \textbf{0.15} & 0.21           \\
			\hline \hline
			\multirow{3}{*}{SIR (dB)}          & Digit & -4.73 & -5.06  & -5.06 & -5.01         & \textbf{3.02}  \\
			& Drums & -6.48 & -5.51  & -1.65 & -5.69         & \textbf{10.21} \\
			& Piano & 0.53  & 2.21   & -3.87 & 2.60          & \textbf{5.12} \\ \hline
		\end{tabular}
	}
\end{table}

\section{Empirical Evaluation}

 
In this section, we evaluate our proposed approach on two source and three source separation experiments on the publicly available Spoken Digit (SC09), drum sounds and piano datasets. The SC09 dataset is a subset of the Speech Commands dataset~\cite{warden2018speech, donahue2018adversarial} 
containing spoken digits (0-9) each of duration $\sim$ 1s at 16kHz from a variety of speakers recorded under different acoustic conditions. The drum sounds dataset~\cite{donahue2018adversarial}
contains 
single drum hit sounds each of duration $\sim$ 1s at 16kHz. The piano dataset~\cite{donahue2018adversarial} contains 
piano music (Bach compositions) each of duration ($>$ 50s) at 48kHz.

\noindent \textbf{WaveGAN Training}. Following \cite{donahue2018adversarial}, we train WaveGAN models on normalized 1s slices (\textit{i.e} $d = $16384 samples) of the SC09 (Digit), Drums and Piano train datasets resampled to 16kHz respectively. All the models were trained using batches of size 128. The generator and discriminator were optimized using the WGAN-GP loss with the Adam optimizer and learning rate 1$e^{-4}$ for 3000 epochs. The trained generator models were used to construct the GAN priors. 

\noindent \textbf{Setup}. For the task of two source separation ($K = 2$), we conducted experiments on three possible mixture combinations: (i) Digit-Piano, (ii) Drums-Piano and (iii) Digit-Drums. In order to create the input mixture for every combination, we randomly sampled (with replacement) normalized 1s audio slices from the respective test datasets, and obtained $1000$ mixtures through a simple additive mixing process. Similarly, we obtained 1000 mixtures for the case of $K=3$, i.e., on the combination, Digit-Drums-Piano. In each case, we performed the PGD optimization using Eqn.\ref{totalloss} for $1000$ iterations with the ADAM optimizer and learning rate  of 5$e^{-2}$ to infer source specific latent features $\{\mathbf{z}^{*}_i\}_{i =1 \dots K}$. The estimated sources are then obtained as $\{ \mathcal{G}_i(\mathbf{z}^{*}_i)\}_{i =1 \dots K}$. Though the choice of initialization for $\mathbf{z}_i$ is known to be critical for PGD optimization~\cite{anirudh2020mimicgan}, we find that setting $\{\mathbf{z}_i\}_{i =1 \dots K} = \mathbf{0} \in \mathbb{R}^{d_z}$ to be effective.  

\noindent \textbf{Evaluation Metrics}. Following standard practice, we used three different metrics - (i) mean spectral SNR~\cite{spiertz2009source, virtanen2007monaural}, a measure of the quality of the spectrogram reconstruction; (ii) mean RMS envelope distance~\cite{morgado2018self} between the estimated and true sources; and (iii) mean signal-interference ratio (SIR)~\cite{SiSEC18} to quantify the interference caused by one estimated source on another.

\noindent \textbf{Results}. Tables \ref{digit_piano}, \ref{drums_piano}, \ref{digit_drums} and \ref{digit_drums_piano} provide a comprehensive comparison of the proposed approach against the standard baselines (FastICA, PCA, KernelPCA, NMF)~\cite{scikit-learn} as well as with the state-of-the-art unsupervised Deep-Audio-Prior~\cite{tian2019deep}. It can be observed that our approach significantly outperforms all the baselines in most cases, except for the Digits-Drums experiment where our method is in par with DAP. These results indicate the effectiveness of our unsupervised approach on complex source separation tasks. We find that the spectral SNR metric, which is relatively less sensitive to phase differences~\cite{defossez2018sing, spiertz2009source}, is consistently high with our approach, indicating high perceptual similarities between estimated and the ground truth audio. We also find lower envelope distance estimates, further emphasizing the perceptual quality of our estimated sources. Finally, we attribute the significant improvements in the SIR metric to the \textit{source dissociation loss} ($\mathcal{L}_{sd}$), which enforces the estimated sources from the priors to be systematically different.




\section{Conclusions}
In summary, we find that source-specific \textit{GAN Priors} are effective in recovering the constituents of an unlabeled mixture, often significantly outperforming unsupervised state-of-the-art benchmarks. Additionally, we find that such generative priors can be further improved with PGD-style optimization using carefully designed spectral domain loss functions. Our approach is highly flexible because it is entirely an inference-time technique, and as a result can efficiently deal with varying number of known sources in a given mixture. This is in contrast with standard supervised approaches which require re-training or extensive fine-tuning. Future extensions to our work include performing source separation when the mixing process is unknown, and dealing with mixtures that contain novel sources.     

\bibliographystyle{IEEEtran}

\bibliography{mybib}


\end{document}